# Two-Level Systems in Nucleated and Non-Nucleated Epitaxial alpha-Tantalum films

Loren D. Alegria[1*], Daniel M. Tennant[1,2], Kevin R. Chaves[1], Jonathan R. I. Lee[1], Sean R. O'Kelley[1],

Yaniv J. Rosen[1], and Jonathan L. DuBois[1]

[1] Lawrence Livermore National Laboratory, Livermore, California 94551, USA

[2] Current address: Rigetti Computing, Berkeley, California 94710, USA

[*]Corresponding Author: alegria4@llnl.gov

## ABSTRACT

Building usefully coherent superconducting quantum processors depends on reducing losses in their constituent materials.[1] Tantalum, like niobium, has proven utility as the primary superconducting layer within highly coherent qubits.[2,3] But, unlike Nb, high temperatures are typically used to stabilize the desirable body-centered-cubic phase, $\alpha$-Ta, during thin film deposition. It has long been known that a thin Nb layer permits the room-temperature nucleation of $\alpha$-Ta,[4-6] but here we observe the epitaxial process and present few-photon microwave loss measurements in Nb-nucleated Ta films. We compare resonators patterned from Ta films grown at high temperature (500 °C) and films nucleated at room temperature, in order to understand the impact of crystalline order on quantum coherence. In both cases, films grew with $Al_2O_3$ (001) || Ta (110) indicating that the epitaxial orientation is independent of temperature and is preserved across the Nb/Ta interface. We use conventional low-power spectroscopy to measure two level system (TLS) loss, as well as an electric-field bias technique to measure the effective dipole moments of TLS in the surfaces of resonators. In our measurements, Nb-nucleated Ta resonators had greater loss tangent ($1.5 \pm 0.1 \times 10^{-5}$) than non-nucleated ($5 \pm 1 \times 10^{-6}$) in approximate proportion to defect densities as characterized by X-ray diffraction (0.27 ° vs 0.18 ° [110] reflection width) and electron microscopy (30 nm vs 70 nm domain size). The dependence of the loss tangent on domain size indicates that the development of more ordered Ta films is likely to lead to improvements in qubit coherence times.[1,7] Moreover, low-temperature $\alpha$-Ta epitaxy may enable the growth of microstate-free heterostructures which would not withstand high temperature processing.[8]

# MANUSCRIPT TEXT

Conventional superconducting qubit microfabrication begins with a superconducting film which is patterned into the gross features of the circuit, providing a low-loss foundation for the preservation of quantum information.[1] Niobium is a popular choice for this layer partly due to broad historical use in other superconducting circuits such as single flux quantum processors.[9] Its critical temperature ($T_C$ = 9.3 K) simplifies refrigeration, and its stability (melting point, $T_m$ = 2750 K) allows consistent device preparation. But for qubits operating at millikelvin temperatures, microwave-frequency loss has become an overarching criterion for building processors with sufficient coherence to entangle more than a few qubits[1]. The stringency of this requirement has drawn qubit fabrication into the domain of surface science, since the electromagnetic fields storing information interact with the interface between the solid state and atmosphere to a degree not usually met with in superconducting or conventional information processing devices[9-12]. Among superconductors being pursued for favorable surface properties[13], tantalum shares desirable traits with niobium but has a distinct surface chemistry[14,15].

Two initial studies exploring qubits with predominantly Ta superconductors have shown coherence times, $T_1$ = 300 - 500 μs, on par with the longest-lived planar qubits made from conventional materials.[3] Although it is tempting to attribute this outstanding performance entirely to the high quality of the tantalum oxide[15], it likely derives in part from the overall technical maturity of those experiments, recommending further systematic comparison of low-power loss in coplanar resonators as a proxy for material performance in a qubit[13]. Recent measurements of Ta resonators with very low loss have corroborated the utility of Ta for qubits, and motivate us to understand quantitatively how loss depends on Ta film structure[16].

Resonator loss is understood within a two level system (TLS) model of defects in the aluminum and niobium surfaces of typical microfabricated superconducting qubits, which commonly consist of niobium on a sapphire substrate with aluminum structures forming the Josephson junction[1,17-19]. Despite extensive study, the origin and physical structure of contributing defects remains uncertain[12]. Nonetheless, the oxides of Nb contribute substantially[12,20], while Ta is known to have a low loss oxide, $Ta_2O_5$.[15,21,22]

It is not clear what the optimal structure would be for reducing TLS loss in qubits. The metal-air interface is usually the most lossy surface type, and yet the strongest fields are present at the edges and substrate-metal interface, so that the optimal metal must form an ordered interface to the substrate as well as to air.[12] This consideration is one motivation for investigating Nb-nucleated films, as Nb is known to form a good interface with sapphire and Ta to air. Moreover, deposition at lower temperatures improves back-end-of-line compatibility with complex microfabrication processes and limits diffusion processes such as silicide formation in the case of fabrication on silicon substrates.[16]

Tantalum stabilizes in a body-centered-cubic structure ($\alpha$-Ta) under equilibrium conditions, but deposition at room temperature typically produces an initial amorphous layer from which surface energetics favors the growth of the metastable $\beta$-Ta phase, a Franck-Kasper phase with a 30-atom unit cell[4,6]. Annealing or growth at elevated temperatures ($\gtrsim$ 500 °C) will form the $\alpha$ phase. The $\alpha$ phase will also grow readily if even a thin crystalline nucleation layer (e.g. 1 nm Nb) is deposited prior to Ta deposition.[5] The more orderly $\alpha$ phase exhibits a higher $T_c$ (4.4 K) and lower resistivity ($\rho \approx$ 20 $\mu\Omega$cm) than the $\beta$ phase (0.5 K, $\rho \approx$ 200 $\mu\Omega$cm)[6,23].

The epitaxy of Ta has been investigated, but almost exclusively at elevated substrate temperatures ($\gtrsim$ 700 °C) typically assumed necessary for high quality growth.[24-26] Given the nearly identical interatomic lattice parameters of bcc Nb and Ta ($\delta$a/a < 0.1 %) their epitaxy on sapphire follow a similar model and heterostructures between Nb and Ta are strain-free.[24,25,27] As with Nb, the c-plane (001) of sapphire supports either (111) or (110) growth modes of Ta. Herein we observe the (110) mode which has three domain types, with (110) aligned and each domain rotated 0 ° or ± 60 ° around (110). This domain structure is illustrated in the inset of Fig. 1a, which shows the oxygen and tantalum coordinates (bulk projections) at the interface between $Al_2O_3$ (001) // Ta (110) for three azimuthally rotated directions. The correspondence of our observations to the known epitaxy of Nb suggests that similar refinements of Ta growth are possible as have been achieved in the case of Nb, and may improve qubit performance.[7,16,24]

To study nucleated Ta for applicability to quantum information-preserving circuits, we grew films by magnetron sputtering on silicon and sapphire wafers and compared these with commercially produced films from the same source as those processed into highly coherent qubits in Ref. [2] Figure 1 summarizes our structural characterization, including films subsequently studied at low temperature. Although three in-house films are presented, they are representative of a larger set of films deposited within our investigation to verify the robustness of the deposition conditions. Further details of the growth and fabrication are included in the Supplemental Information.

X-ray diffraction (XRD) in Fig. 1a shows that $\alpha$-Ta was formed with the introduction of a 2 nm Nb layer. We compare four wafers, each containing a 200 nm thick Ta layer: Sample 1 was grown commercially on sapphire at 500 °C to stabilize the $\alpha$ phase, as confirmed by the strong $\alpha$-Ta (110) reflection and absence of the $\beta$-Ta (002) reflection. Sample 2 was grown in-house at room temperature with a 2 nm Nb nucleation step and shows a qualitatively identical spectrum to the non-nucleated growth. Sample 3 also contained a nucleation step, but on a silicon (100) substrate, leading to an $\alpha$-Ta phase, but with a less intense (110) peak and none of the additional peaks (30 ° < $2\theta$ < 35 °) seen in Samples 1 – 2. Finally, Sample 4 was grown on silicon with no nucleation and at 400 °C, producing a mixed phase, the proportions of which are consistent with deposition-temperature studies in the literature (c.f. [6,16]). From these data it is clear that including 2 nm Nb precipitates the $\alpha$ phase, but the additional peaks of Samples 1 – 2 point to substantial structural differences between growth on sapphire and on Si.

High resolution scanning electron microscopy (SEM) (Fig. 1b-d) clarifies the film structures. On sapphire, both nucleated and non-nucleated films contain elongated ridges either oriented in parallel or rotated ± 60 °. The ridges are aligned across the entire sapphire wafer, confirming the presence of long-range order and an epitaxial relationship between the sapphire and Ta. By contrast, the $\alpha$ film on Si (100) has similar domains which are oriented randomly (Fig. 1d), and $\beta$ phase films are featureless or have ~ 20 nm hillocks (e.g. Fig. 1e). The Fourier transforms of the SEM images (Fig. 1b-e, right) show the domain size distribution in each case, and a clear alignment of the Ta with the hexagonal basal plane of the sapphire. The $Al_2O_3$ (001) // Ta (110) growth mode is consistent with prior epitaxial studies of Ta grown at high temperature under conditions of relatively high contaminant flux, suggesting that qubit devices using (111)-oriented Ta may achieve even better performance than the aforementioned qubit measurements.[2,24] Note that the imaging here is performed with a 45 ° substrate tilt to enhance contrast. In the case of the nucleated film, the preservation of the crystallographic orientation across the Nb/Ta boundary testifies to the low strain of that heterostructure.

The additional XRD peaks of Samples 1 and 2 (30 ° < $2\theta$ < 35 °) appear to arise from the strain fields particular to the domain structure. These peaks were observed in Ref. [2] and ascribed to contamination or instrumental artifacts, but our observations confirm them to be a common feature of $\alpha$-Ta films on sapphire. Their absence in the Ta on Si implies they do not arise from oxides, and their widths correlate with the other Ta features[6].

In the following we turn to low-temperature measurements of the quantum coherence properties of the non-nucleated and the Nb-nucleated α-Ta films, Samples 1 and 2. To summarize the structural characterization: Sample 1 exhibits larger mean crystal grain size, sharper (110) XRD peaks, and lower room-temperature resistivity (70 nm, 0.18 °, 22 µΩcm, $T_C$ = 4.3 K, RRR = 7.8) than the nucleated Sample 2 (30 nm, 0.27 °, 24 µΩcm, $T_C$ = 4.1 K, RRR = 2.6) indicating a larger density of defects in the nucleated film. Here RRR is equal to the ratio between the room temperature and 4.2 K resistivity. Given that the grain boundary area is inversely proportional to the grain size, and defects are concentrated on the grain boundaries, we can estimate the defect density to be two times greater in the nucleated sample.

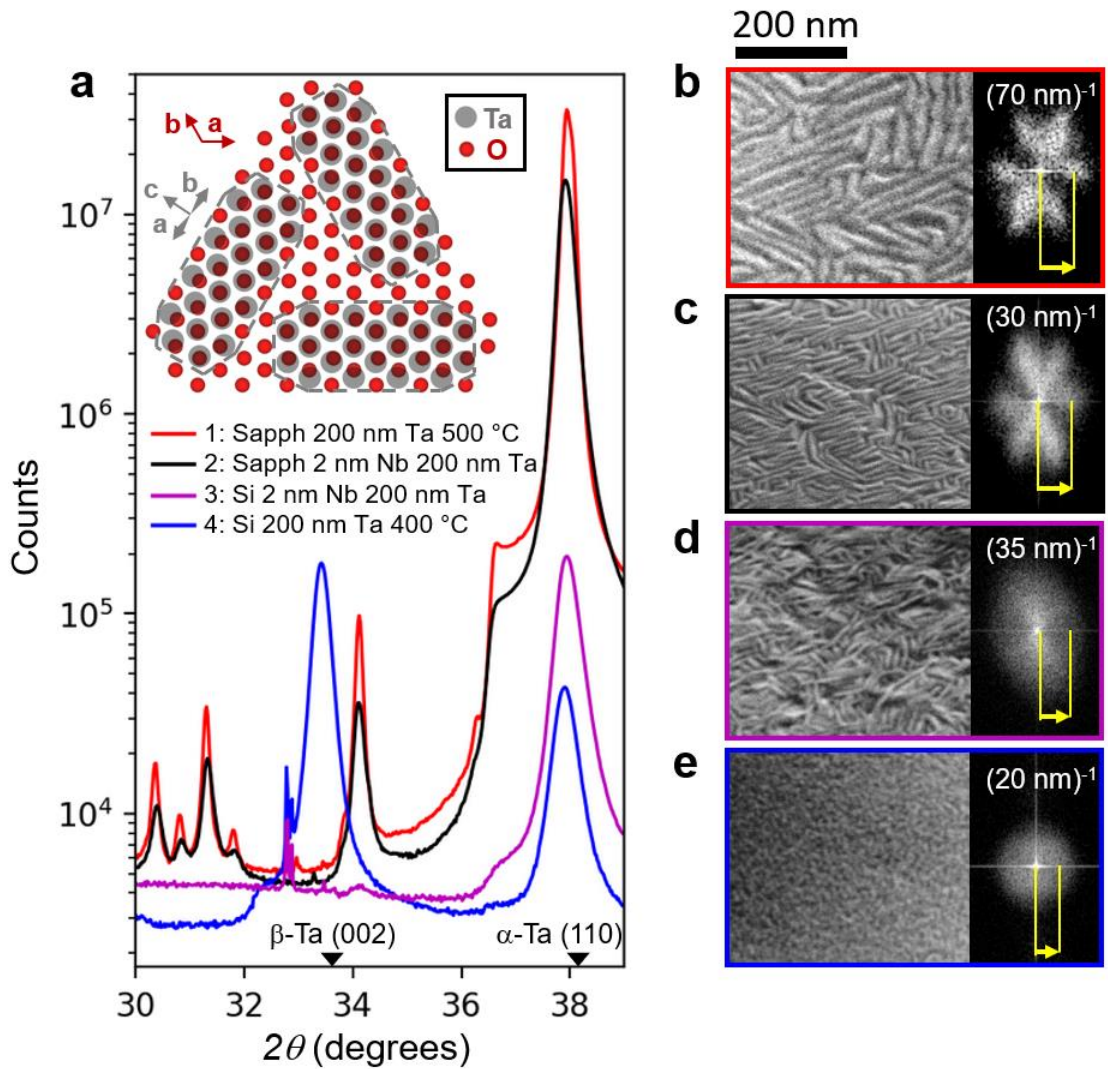

**Figure 1.** Nucleation and room temperature epitaxy of α-Ta on sapphire. **A** X-ray diffraction data in the region 2θ = 30 – 40 ° allows for detection of the α and β phases by the presence of their (110) (2θ

= 38.5 °) and (002) (2θ = 33.7 °) lines, respectively. The four spectra here represent (1) an α-Ta film grown on c-plane sapphire at 500 °C, (2) an α-Ta film grown on c-plane sapphire by nucleation at room temperature, (3) an α-Ta film nucleated on silicon, (4) a Ta film grown on silicon at 400 °C without a nucleation step, yielding a mixed α/β phase. (inset) The three orientations in which interfacial Ta and O align in $Al_2O_3$ (001) // α-Ta (110) are consistent with the three domain orientations observed in SEM. **B** In SEM, the α-Ta films on sapphire had a hexagonally faceted topography, with facets aligned to the hexagonal surface presented by the sapphire, indicating epitaxial growth.  The Fourier transform of the SEM images (right) allows us to measure the mean grain size and confirm the orientation relative to the wafer facets.  **C** In the nucleated film on sapphire, the film orientation was preserved but with reduced grain size. **D** In the nucleated film on silicon, a similar grain size is evident in SEM, but there is no registry between the grains and the substrate. **E** The mixed phase sample showed no elongated grain growth.

Samples 1 and 2 were patterned into hanger-type resonators (Fig. 2) following the design of Ref. [18]. In this geometry, a low-frequency (< 40 kHz) voltage bias can be applied to the resonator surface, and the bias line is isolated from the resonators and coplanar waveguide by > 50 dB, permitting TLS-limited loss to be measured in each resonator while gating the dielectrics with an electric field.

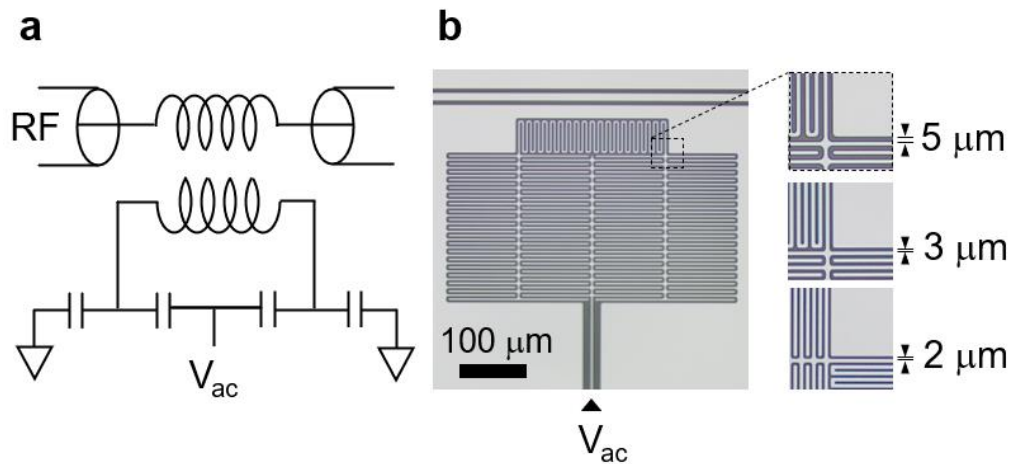

**Figure 2:** Biased Resonator Geometry. **(a)** Samples 1 and 2 were patterned into resonators and characterized by transmission (S21) through a coplanar waveguide (RF) to which they are inductively coupled. An interdigitated bias electrode ($V_{ac}$) is decoupled from the resonator at the resonance frequencies (6 - 8 GHz), but permits a low-frequency (kHz) electric field to be applied to the surface of the resonator for TLS structure characterization. **(b)** Optical inspection showed the resonators to be free from major defects. The three resonator designs measured differed by their dielectric gap distance.

We conducted conventional loss measurements in a dilution refrigerator as summarized in Figure 3. Resonator lineshapes (examples in Fig. 3a,c) fit the asymmetrical Lorentzian model[28], from which we obtain the internal quality factors for a given photon occupation number, $n_p = 2PQ_L^2/(\hbar Q_c (2\pi f_0)^2)$, where $P$ is the incident power at the sample, $Q_L$ is the resonator loaded quality factor, $Q_c$ is the coupling quality factor, and $f_0$ is the resonator frequency. From these we plot the loss tangent, $tan(\delta) = 1/Q_i$ in Fig. 3b,d as a function of $n_p$. The high-power loss tangent is proportional to $n_P^{-0.12}$ similar to previous studies of microfabricated resonators.[13,29] From the fits we find the average low-power loss tangent across the three geometries to be $1.5 \pm 0.1 \times 10^{-5}$ for the nucleated and $5 \pm 1 \times 10^{-6}$ for the non-nucleated samples. Various studies have found that device and enclosure design can strongly impact observed loss tangents, and while we measured each sample using identical boxes and lithographic patterns, these may introduce different loss anomalies as compared to other studies.[13,16] If we assume equivalent TLS dipole moment, $p$, the density of TLS ($\rho \sim tan(\delta)/p$) in the nucleated samples is three times greater, similar to the estimated relative density of defects. Although the theoretical relationship between the defect density in a metal and in its oxide, where TLS are presumed to form, is far from obvious, our findings corroborate recent evidence that grain size affects low power losses in these materials[7].

.

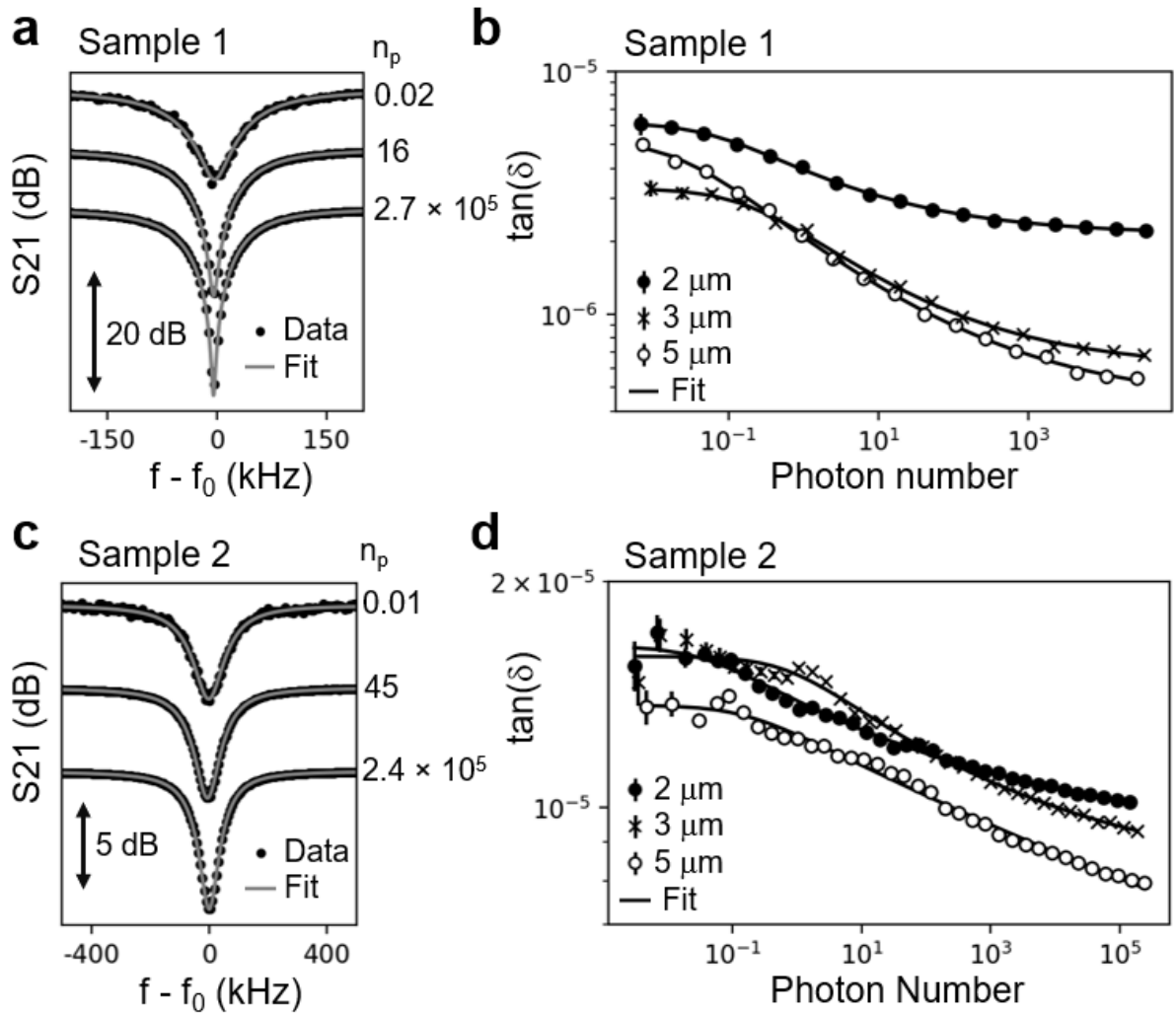

**Figure 3:** TLS loss measurements in non-nucleated and nucleated films (Samples 1 and 2 respectively). **(a)** Raw S21 data were fit by an asymmetric Lorentzian model for each value of incident microwave power (equivalently photon number, $n_{ph}$). **(b)** The loss tangent, $\tan(\delta)$ for each resonator design is dominated by TLS in the few-photon limit, and is accurately fit by a model of TLS saturation at high power. **(c)-(d)** Measurements in resonators patterned from nucleated Ta showed three times greater loss than non-nucleated.

Next we measured the loss tangent as a function of voltage bias frequency to measure the dipole moment of TLS in the device surfaces. Following the method of Ref. [18], we apply a low frequency triangle wave (100 - 40 kHz) with fixed peak-to-peak amplitude ($V_{ac}$ = 5 V) to the resonator bias electrode. The changing electric field, $E_{ac}$, has the effect of bringing defect energy levels in and out of resonance with the resonator mode. As the resonator overlaps with more defect levels, more energy is transferred out of the resonator. Further, the greater the polarizability of the TLSs, the more TLS energy levels the bias brings into degeneracy with the resonator. Consequently, the change in loss observed with change in bias frequency allows for measurement of the mean TLS dipole moment.

Quantitatively, a Landau-Zener model describes the interaction between a time-varying TLS mode and the resonator mode[18]. The two interact with a Rabi frequency $\Omega_r = pE_{res}\Delta_0\cos(\theta)/\hbar\omega$ where $p$ is the TLS dipole moment, $E_{res}$ is the resonator electric field, $\theta$ is the angle between the dipole moment and the electric field, $\Delta_0$ is the TLS tunneling energy, and $\omega$ is the resonator frequency. In the changing bias field $E_B(t)$, the TLS energy level will change at a rate $\omega' \equiv d\omega/dt = 2p\cos(\theta) E_B'(1-(\Delta_0/\hbar\omega)^2)^{1/2}/\hbar$ and Landau-Zener theory predicts an energy exchange with probability $P = 1-\exp(-\pi\Omega_r^2/2\omega')$. A time averaged local loss tangent can be calculated as a function of $E_B'$ by integrating over a logarithmic distribution of tunneling energies and uniformly distributed dipole orientation in space, with the result[17]

$$\tan\delta_{local} = 3\xi \tan(\delta_{mat}) \int_0^1 d\alpha \int_0^1 d\beta \frac{\beta}{\alpha}\left(1 - \exp\left(-\frac{\alpha^2\beta}{\xi\sqrt{1-\alpha^2}}\right)\right), \qquad (1)$$

where $\xi = 4E_B' \hbar/\pi pE_{res}^2$ is the local Landau-Zener parameter and $tan(\delta_{mat})$ is the static loss tangent for a given material (substrate or oxide). To account for the spatial variations of electric field, a finite-element model of the resonator is constructed, from which we calculate $E'_B(r)$ and $E_{res}(r)$ as a function of $r$, the location within the resonator. The local loss tangent (Eq. 1) is then integrated over volume, scaling according to the local field participation, to obtain the overall loss tangent of the resonator: $\tan\delta = \int dV \frac{1}{2}\epsilon E_{res}^2 \tan\delta_{local} / \int dV \frac{1}{2}\epsilon E_{res}^2$. As a function of $\xi$, the loss tangent follows the characteristic dependence given by Eq. 1 independent of the frequency, to which we compare the experimental results and calculate the dipole moment.

In Figure 4, we present the results of bias frequency sweeps in resonators with nominally identical geometries (5 μm gap) from Sample 1 (non-nucleated Ta) and Sample 2 (nucleated Ta). The voltage bias shifts the loss-tangent (Fig. 4a,c) in a manner consistent with the model described above. This can be seen by plotting the data as a function of $f_{ac}/n_P \sim \xi$. Doing so, the data from the various frequencies all fall upon a single curve. The only unknown factor in $\xi$ is $p$, the effective dipole moment, and performing a regression on this value to fit the theoretical model, we obtain $p$ = 60 ± 3 Debye and 10.7 ± 0.7 Debye for the non-nucleated and nucleated samples, respectively (Fig. 4b,d). The dipole moments are larger than typically observed[17,19,30], but relatively few studies have endeavored to quantify the dipole moments in the surface oxides, and anomalously large moments may result naturally from the particular energy landscape of the interfaces involved[31]. One source of uncertainty in calculating the dipole magnitude stems from the possibility of a non-uniform distribution of TLS within the device, so that the ratio between the moments for the two identical patterns is more reliable.

Although the advantage of Ta for quantum circuits is its simplicity—its presumably homogeneous oxide as compared to Nb—our study has demonstrated that processing differences can strongly alter film structure and loss. Given prior literature on Nb growth, it appears there remains space for improvement of Ta growth for quantum processors. Moreover, our findings indicate that this may be accomplished even at moderate temperatures, allowing for structures which would not be stable at higher temperatures. Finally, the disparate TLS dipole moments we observe suggest that grain structure may influence electronic structure at the metal surface, and underscore the importance of developing surface metrology methods specific to quantum materials development.

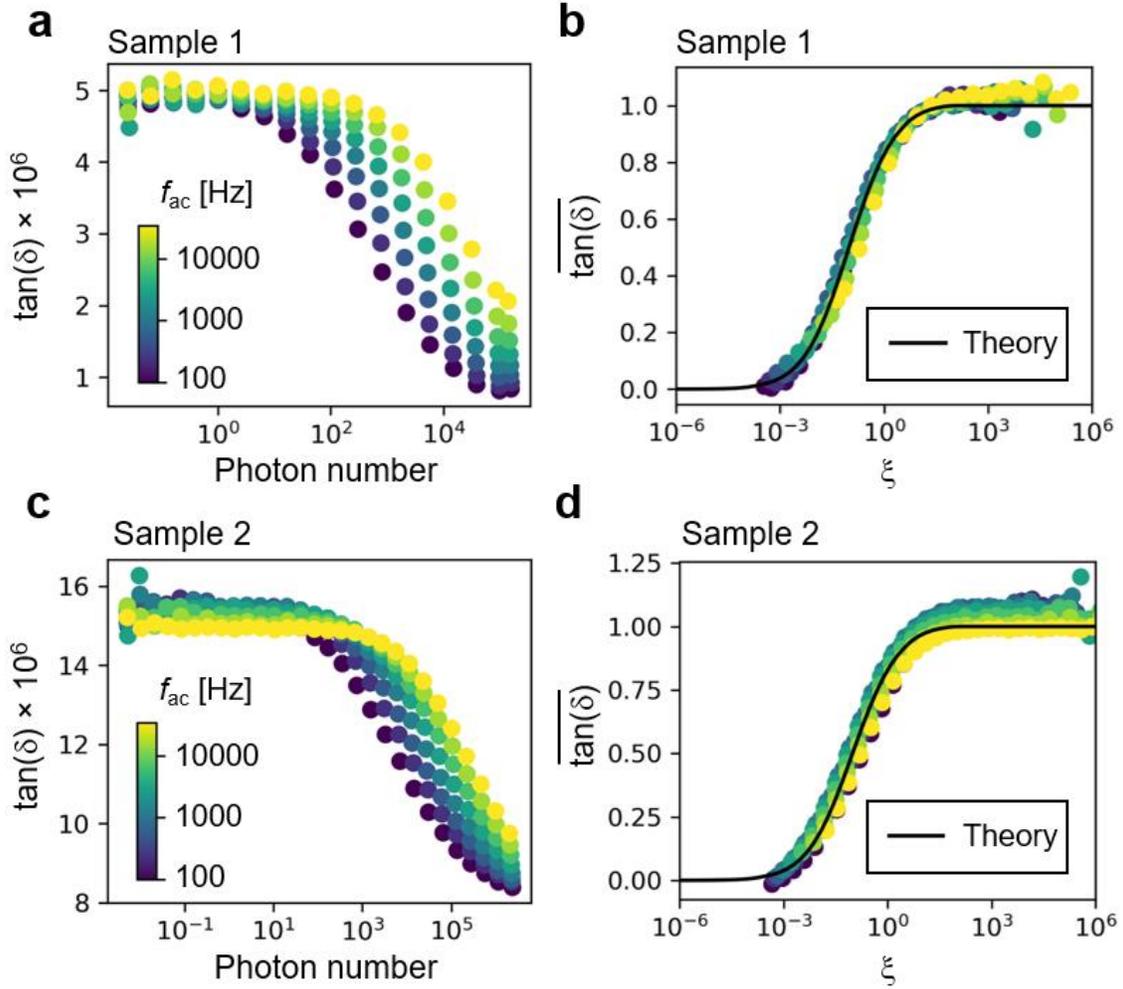

**Figure 4:** Applying a variable-frequency drive to the bias electrode induces a shift in the loss tangent curve consistent with Landau-Zener energy exchange between the resonator and the TLS population. The model implies an effective TLS dipole magnitude in a given resonator. **(a),(c)** Bias drive-induced loss tangent shift for a non-nucleated and a nucleated sample. **(b),(d)** Plotting the loss versus the dimensionless Landau-Zener parameter $\xi = 4E_B' \hbar/\pi p E_{res}^2$ (see text) the data fall upon a universal curve which determines $p$, the effective TLS dipole moment, which is found to be six times greater in the non-nucleated sample as compared to the nucleated sample.

# AUTHOR DECLARATIONS

## Conflict of Interest

The authors have no conflict of interest to disclose.

## Author Contributions

**Loren D. Alegria**: Investigation (lead), Conceptualization (equal); Data Curation (equal); Formal analysis (equal); Visualization (equal); Writing – original draft (lead); Writing – review & editing (equal). **Daniel M. Tennant:** Investigation (supporting); Writing – review & editing (equal); Formal analysis (equal); Data Curation (equal). **Kevin R. Chaves:** Investigation (supporting); Writing – review & editing (equal). **Jonathan R. I. Lee:** Investigation (supporting) ; Writing – review & editing (equal). **Sean R. O'Kelley:** Investigation (supporting); Writing – review & editing (equal). **Yaniv J. Rosen:** Conceptualization (equal); Investigation (supporting); Writing – review & editing (equal); Formal analysis (equal). **Jonathan L. DuBois:** Conceptualization (equal); Writing – review & editing (equal).

# SUPPLEMENTARY MATERIAL

Film deposition conditions, XRD measurement conditions, resonator fabrication protocols and cryogenic measurement instrumentation are detailed in the supplementary material.

# ACKNOWLEDGEMENTS

This work was supported by LLNL-LDRD-22-FS-040 (supporting the contributions of LDA, KC, and JRIL) and by DOE SC BES under Award DE-SC0020313 (supporting the contributions of YJR, SRK, JLD, and DMT) under the auspices of Lawrence Livermore National Laboratory (LLNL) under Contract No. DE-AC52-07NA27344.

# DATA AVAILABILITY

Data that support the findings of this study are available within the article and additional data that support the findings are available from the corresponding authors upon reasonable request.